%Paper: hep-th/9305125
%From: "David Lowe" <lowe@puhep1.Princeton.EDU>
%Date: Mon, 24 May 93 17:18:15 EDT

\input harvmac

%%%%%%%%%%%%%%%%%%%%%%%%%%%%%%%%%%%%%%%%%%%%%%%%%%%%%%%%%%%%%%%
%The following lines are needed to insert the accompanying figures in
%the paper. If you do not have epsf, then comment out the line
% ``\input epsf'', and print the figures separately. The figures are at
%the end of the tex file, with instructions for their extraction.
\input epsf
\ifx\epsfbox\UnDeFiNeD\message{(NO epsf.tex, FIGURES WILL BE IGNORED)}
\def\figin#1{\vskip2in}% blank space instead
\else\message{(FIGURES WILL BE INCLUDED)}\def\figin#1{#1}\fi
\def\ifig#1#2#3{\xdef#1{fig.~\the\figno}
\goodbreak\topinsert\figin{\centerline{#3}}%
\smallskip\centerline{\vbox{\baselineskip12pt
\advance\hsize by -1truein\noindent{\bf Fig.~\the\figno:} #2}}
\bigskip\endinsert\global\advance\figno by1}
%%%%%%%%%%%%%%%%%%%%%%%%%%%%%%%%%%%%%%%%%%%%%%%%%%%%%%%%%%%%%%%%

\def\scri{ {\cal I}}
\Title{\vbox{\baselineskip12pt\hbox{hep-th/9305125}\hbox{RU-93-12}
\hbox{PUPT-1399}\hbox{NSF-ITP-93-56}}}
{\vbox{\hbox{\centerline{Nonsingular Black Hole Evaporation and
``Stable'' Remnants}}}}
\centerline{\it D.A. Lowe\foot{lowe@puhep1.princeton.edu}}
\centerline{Department of Physics}
\centerline{Princeton University}
\centerline{Princeton, NJ 08544}
\smallskip
\centerline{\it and}
\smallskip
\centerline{\it M. O'Loughlin\foot{ologhlin@physics.rutgers.edu}}
\smallskip
\centerline{Department of Physics and Astronomy}
\centerline{Rutgers University}
\centerline{Piscataway, NJ 08855-0849}
\noindent
\smallskip
\vskip .5cm
\baselineskip 22pt
\noindent
\vbox{ \baselineskip12pt
We examine the evaporation of two--dimensional black holes, the classical
space--times of which are extended geometries, like for example the
two--dimensional section of the extremal
Reissner--Nordstrom black hole. We find that the
evaporation in two particular models proceeds to a stable end--point.
This should represent the generic behavior of a certain class
of two--dimensional dilaton--gravity models. There are two
distinct regimes depending on whether the back--reaction is weak
or strong in a certain sense.
When the back--reaction is weak, evaporation proceeds via an adiabatic
evolution, whereas for strong back--reaction, the decay proceeds
in a somewhat surprising manner. Although
information loss is inevitable in these models
at the semi--classical level, it is rather benign, in that the
information is stored in another asymptotic region.}

\Date{May, 1993}
%\draftmode

\lref\GMGHS{G. Gibbons and K. Maeda, {\it Nucl.
Phys.}{\bf B298}(1988), 741; D. Garfinkle, G. Horowitz, and A.
Strominger, {\it Phys. Rev.}{\bf D43},(1991),3140; Erratum: {\it Phys. Rev.}
{\bf D45} (1992), 3888.}
\lref\bddo{T. Banks, A. Dabholkar, M.R. Douglas, and M. O'Loughlin, ``Are
horned particles the climax of Hawking evaporation?'' {\it Phys. Rev.}
{\bf D45}(1992),3607.}
\lref\corn{S.B. Giddings and A. Strominger, ``Dynamics of
extremal black holes'', {\it Phys. Rev.} {\bf D46} (1992) 627;
M.Alford,A.Strominger, ``S Wave Scattering of Charged
Fermions by a Magnetic Black Hole'', {\it Phys.Rev.Lett.}{\bf
69},(1992),563.}
\lref\bol{T. Banks and M. O'Loughlin,
``Classical and quantum production of cornucopions at energies below
$10^{18}$ GeV'', Rutgers preprint RU-92-14 (1992).}
\lref\gidstrom{S. Giddings and A. Strominger, ``Dynamics of extremal
black holes'',{\it Phys. Rev.}{\bf D46} (1992), 627.}
\lref\bghs{B. Birnir, S. Giddings,
J. Harvey, A. Strominger, ``Quantum Black Holes''
{\it Phys. Rev. }{\bf D46}(1992), 638.}
\lref\hawk{S. Hawking, ``Evaporation of two-dimensional black holes'',
{\it Phys. Rev. Lett.}{\bf 69}(1992), 406.}
\lref\sthandb{L. Susskind, L. Thorlacius,
``Hawking Radiation and Back-Reaction'', {\it Nucl. Phys. }{\bf
B382}(1992), 123.}
\lref\hastew{S.W. Hawking, J.M. Stewart, ``Naked and Thunderbolt
singularities in Black Hole Evaporation'' Cambridge preprint,
PRINT-92-0362, hep-th/9207105 (1992).}
\lref\dlowe{D.A. Lowe, ``Semiclassical approach to Black Hole
Evaporation,'' {\it Phys. Rev.} {\bf D47} (1993), 2446.}
\lref\rst{J. Russo, L. Susskind, and L. Thorlacius, ``
Black hole evaporation in 1+1 dimensions'' {\it Phys. Lett.}{\bf B292}(1992),
13.}
\lref\cghs{C.G. Callan, S.B. Giddings, J.A. Harvey, and A.
Strominger, ``Evanescent black holes,''  {\it Phys.
Rev.}{\bf D45} (1992) R1005; J. Harvey and A. Strominger, ``Quantum
Aspects of Black Holes'', Enrico Fermi Institute preprint EFI-92-41,
hep-th/9209055, and references cited
therein.}
\lref\nsbh{T.Banks and M.O'Loughlin, ``Nonsingular Lagrangians for Two
Dimensional Black Holes'', hep-th/9212136.}
\lref\triv{S. Trivedi, ``Semiclassical Extremal Black Holes'', Caltech
preprint, CALT-68-1833, hep-th/9211011.}
\lref\sttriv{A. Strominger, S. Trivedi, ``Information
Consumption by Reissner Nordstrom Black Holes'', ITP preprint,
NSF-ITP-93-15, hep-th/9302080.}
\lref\pist{T. Piran, A. Strominger, ``Numerical Analysis of
Black Hole Evaporation'', ITP preprint, NSF-ITP-93-36, hep-th/9304148.}
\lref\bilal{A. Bilal, ``Positive energy theorem and
supersymmetry in exactly soluble quantum corrected 2-d dilaton
gravity'', Princeton preprint, PUPT-1373, hep-th/9301021.}
\lref\bilkog{A. Bilal, I. Kogan, ``Hamiltonian approach to
2-d dilaton gravities and invariant ADM mass'', Princeton preprint,
PUPT-1379, hep-th/9301119.}
\lref\info{T.Banks, M.O'Loughlin and A.Strominger, ``Black Hole Remnants
and the Information Puzzle'',hep-th/9211030,{\it Phys.Rev.}{\bf
D47}(1993),4476.}
\lref\boone{T.Banks and M.O'Loughlin, ``Two--dimensional quantum gravity
in Minkowski space'', {\bf NPB362}(1991),649.}
\lref\polylc{A.M.Polyakov, ``Quantum gravity in two dimensions'',{\it
Mod.Phys.Lett.}{\bf A2}(1987),893.}
\lref\parkst{Y-C.Park and A.Strominger, ``Supersymmetry and positive
energy in classical and quantum two-dimensional dilaton gravity'',{\it
Phys.Rev.}{\bf D47}(1993),1569.}
\lref\dealwis{S.P.de Alwis, ``Two-dimensional quantum dilaton gravity
and the positivity of energy'', hep-th/9302144, COLO-HEP-309.}
\lref\chhart{S.Chandrasekhar and J.Hartle, {\it Proc.R.Soc.London}{\bf
A284}(1982),301.}
\lref\PI{E.Poisson and W.Israel, {\it Phys.Rev.Lett.}{\bf
63} (1989),1663; {\it Phys.Lett.}{\bf B233} (1989),74; {\it Phys.Rev.}{\bf
D41}(1990),1796.}
\lref\ori{A.Ori, {\it Phys.Rev.Lett.}{\bf 67}(1992),789.}

\newsec{\bf Introduction}

One of the authors (M.O'L.) and
T.Banks \ref\nsbh{T.Banks and M.O'Loughlin, ``Nonsingular Lagrangians
for two-dimensional black holes'', RU-92-61, hep-th/9212136.} had
previously proposed that for a large
class of modified scalar-gravity theories in which the classical
geometries are all nonsingular, with causal structure identical to that of
Reissner--Nordstrom, the Hawking evaporation to a final
zero temperature remnant--like
object could be studied without singularity, as opposed to the
original CGHS (Callan-Giddings-Harvey-Strominger) \cghs\
models in which a now well known singularity was found.

Here we report on  calculations that support this picture.
When in--falling matter perturbs one of these extremal solutions,
two apparent horizons form. As the evaporation takes place, these
apparent horizons approach each other.
We find two distinct regimes, depending on whether the
back--reaction is weak or strong in a certain sense.
With weak back--reaction, an adiabatic approximation
gives a correct description, and things settle back down
to a stable remnant, with the apparent horizons
meeting only after an  {\it infinite} proper time.
In the strong back--reaction regime, the apparent horizons
meet after a {\it finite} proper time, and
only after meeting do things settle back down to the extremal solution.
Black holes in these models therefore evaporate in a completely
nonsingular fashion, realizing the original objectives of CGHS.
Information loss occurs at the
semi--classical level, but only in a rather benign way.

In section 2 we introduce the models of interest. We discuss in some
detail the behavior near the double horizon of the extremal static
semi--classical space--time in section 3. In section 4 we describe the
adiabatic approximation \sttriv\ for the nonsingular models. Section 5
contains the results of our numerical analysis and section 6 is devoted
to our conclusions and a discussion of their implications.

\newsec{\bf The Models}

Consider a Lagrangian taken from the general class of
two--dimensional renormalizable generally covariant field theories \boone ,
\eqn\genlag{
{\cal L}_{cl} = \sqrt{-g}(D(\phi) R + G(\phi)(\nabla\phi)^2 + H(\phi))~.}
We require that the potentials behave asymptotically like those of
linear dilaton gravity \cghs ,
\eqn\asymps{
D(\phi) \to {G(\phi) \over  4} \to {{H(\phi)}
\over 4} \to e^{-2\phi}~,}
as $\phi \to -\infty$.

%The two-dimensional target space of this model is Minkowskian.
%, and in
%conformal gauge the $\rho$ direction is light-like.
The renormalization
group equations are hyperbolic on
the two--dimensional target space of this model,
and thus given a set of
initial data one can consistently renormalize the
model \ref\Banklykk{T.Banks and J.Lykken, ``String Theory and
Two--dimensional Quantum Gravity'', {\it Nucl.Phys.}{\bf B331}(1990),173.}.
In the following, without loss of generality, we will restrict attention
to the class of models satisfying $G(\phi) = -2 D^{\prime}(\phi)$.
Other models may be obtained by a field redefinition of $\phi$.
Performing a Brans--Dicke transformation on the metric
${\hat g} = e^{-2 \phi} g$ this Lagrangian
may be rewritten in the simple form
\eqn\simp{
\hat {\cal L}_{cl} = \sqrt{-{\hat g}}(D(\phi) {\hat R} +
W(\phi) )
}
where we have defined $W(\phi) = e^{2 \phi} H(\phi) $. This
form of the Lagrangian is convenient for finding the classical
solutions as described in
\nsbh\ , in which reference the extended
space--time geometries are discussed.

\ifig\fone{Causal structure of the extremal zero temperature space--time
geometries. The arrows show the flow of time.  CD is
the Cauchy horizon for regions I and III. The global horizons are at
fixed $r$ and the geometry near $r = r_{h}$ is the same in both the
DW and
RN cases
(in particular the distance to P in the direction of the arrow is
infinite).  The line ST is singular and at finite spacelike distance for
RN, nonsingular at infinite spacelike distance for DW. }
{\epsfysize=3in\epsfbox{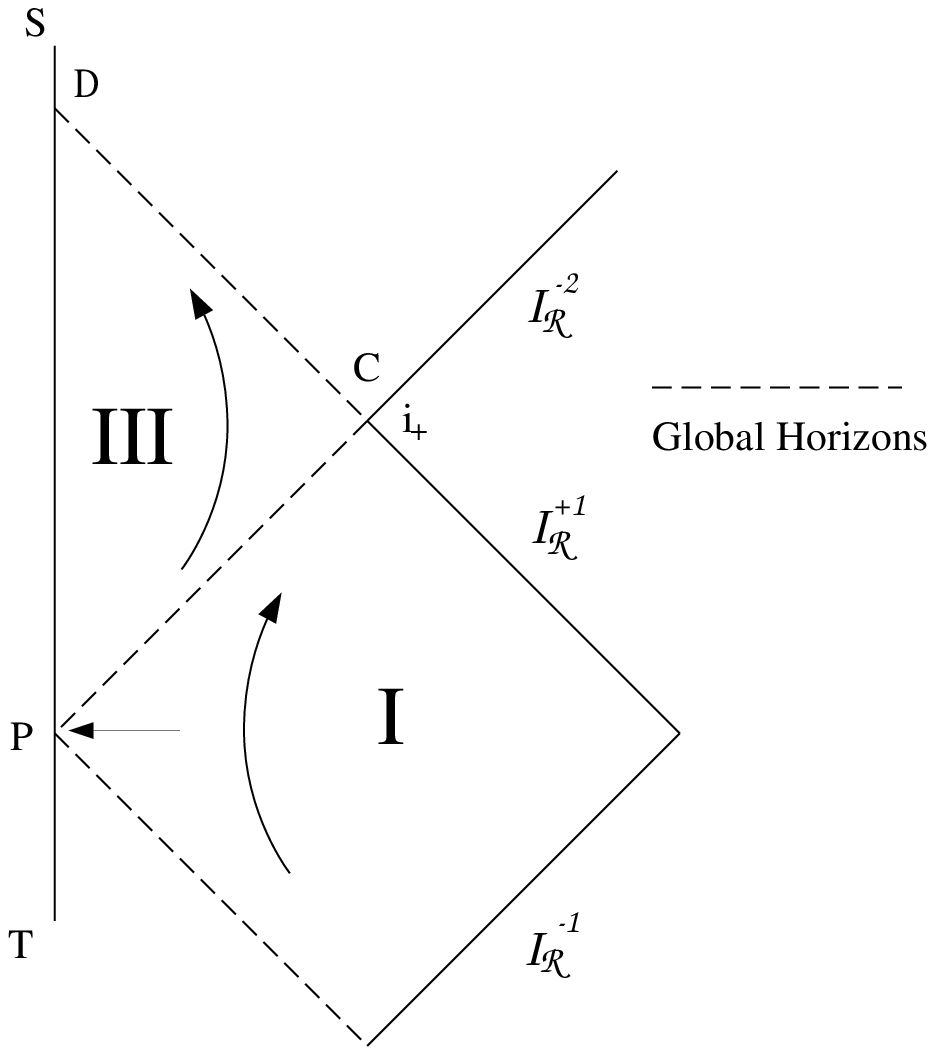}}

All solutions are causally related to the two--dimensional $r-t$ section of
the four--dimensional Reissner--Nordstrom (RN) black hole
(see \fone\ for the extremal geometry),
but as noted above, the geometries in
these generalized models may be nonsingular. The nonsingularity is achieved
by requiring that as $\phi \to \infty$, $D \sim e^{n\phi}$ and $W \sim
e^{m\phi}$, such that $n\geq m-2$.
The multiple horizon structure is obtained by requiring that W have a zero.
Henceforth this class of models will be referred to as DW models.

Just as in four--dimensions one believes that if charge cannot be radiated
away then the zero temperature extremal RN space--time with
$M^2 = Q^2$, where $M$ is the mass and $Q$ is the charge,
will be an end--point for Hawking evaporation, it
was conjectured that the zero temperature extremal limit of the DW
models would also be a natural end--point for Hawking evaporation.

To compute the back--reaction we look at the trace anomaly produced by the
quantum fluctuations of some conformally coupled (hence massless) matter
fields. This is just the quantity calculated by Polyakov. We follow
CGHS and study the back--reaction corrected equations in the
semi--classical limit, obtained by
adding to the equations the terms corresponding to the trace anomaly
of a large number $N$ of the matter fields $f_i$
and then taking $N$ to
infinity with $N\hbar$ fixed. Explicitly, the semi--classical
action is
\eqn\acton{
S= \int d^2x~ {\cal L}_{cl} -2 \kappa \del_+ \rho \del_- \rho
+\sum_{i=1}^N  \del_+ f_i \del_- f_i~,}
in conformal gauge ( $g_{\pm \pm} =0$, $g_{\pm \mp}= -\half e^{
2\rho} $, $x^{\pm} = x^0 \pm x^1$).
Here we have defined $\kappa = N\hbar /12$.

By our choice of potentials, together with the condition
$D' + 2\kappa < 0$ for all
$\phi$, we have avoided the  singular kinetic term in field space
which has been shown to be responsible for the singularity of the CGHS model.

Let us note in passing that the general solution to the CGHS equations that
is static, with mass above that of the vacuum, actually has a weak
coupling singularity on the outer horizon \refs{\hawk,\bghs}\ (the quantum
kink). This singular behavior arises through the interaction between
the assumed dilaton--gravity asymptotics of the potentials \asymps ,
and the large $N$ corrections.
That is, this singularity is a weak coupling singularity that arises
due to the assumed linear dilaton asymptotics of the Lagrangian, and so
also will be present in our models.

Note that we may alternatively require the
potentials to behave asymptotically as the spherically symmetric
reduction of the four--dimensional Einstein equations,
\eqn\ein{
D(\phi) \to {G(\phi) \over  2} \to  e^{-2\phi}, \quad {\rm and} \quad
H(\phi) \to 2 ~.
}
Here the two--dimensional metric is related to the four--
dimensional metric via
\eqn\fmet{
^{(4)}ds^2 =  ds^2 + e^{-2 \phi} d\Omega^2~.
}
An example from this class of models, corresponding to the
$r-t$ section of the four-dimensional RN black hole will also
be considered in the following,
with the same one--loop back--reaction term included
as above. This model has recently been
considered in \refs{ \sttriv,\triv }. Unlike the DW models described
above, this model is singular at the classical level and  has a
large $N$ singularity at $e^{-2 \phi} = \kappa/2 $.

\newsec{\bf Static Solutions of the Generalized Models}

The static solutions to the quantum equations
of the DW models fall into three categories,
those that qualitatively have the same structure as the quantum kink of
early studies of the CGHS Lagrangian, an extremal state that approaches the
classical extremal state (as $\kappa\to 0$),
but with a singularity  in the second
derivative of the
curvature on the horizon, and negative mass solutions that have the same
structure as the classical negative mass solutions, in particular they are
nonsingular. In the following we will refer to the
extremal state as the vacuum of the
two--dimensional theory, and measure mass relative to this
vacuum state.

Thus at this level we have found that quantum corrections have made
the classical phase space more singular. However, the vacuum state is
nonsingular and given that the positive mass solutions will radiate, it is
plausible that none of the positive mass singularities will appear in the
dynamical collapse and evaporation.

In the next section we will introduce the adiabatic approximation used by
Strominger and Trivedi to study the formally similar problem for a
dimensional reduction of Reissner--Nordstrom geometry.
The adiabatic approximation can be
most easily investigated in light--cone gauge, so we now record the
equations in that gauge.

We have the line element,
\eqn\linee{
ds^2 = -h {dv}^2 + 2 dr dv~.
}
The scalar curvature is given by $R = -\partial_r^2 h$. To find the
needed parts of the stress tensor, we integrated the Bianchi identities
$\nabla^\alpha T_{\alpha \beta} = 0$, using the equation for the trace
anomaly \polylc ,
\eqn\anom{
T_\alpha^{\alpha m} = 2T_{rv}^m + h T_{rr}^m = {{\kappa} \over 4} R~.}
The components of interest to us are the trace, and
${\it l}^\alpha {\it l}^\beta T_{\alpha \beta}$, where
${\it l} = (h/2,1)$.

The $\phi$ and trace equations may be written as,
\eqn\lcI{
\nabla^2 \phi = {(e^{-2\phi}(D'W + \kappa(W - \half W')) - D''(D' +
\kappa) (\nabla \phi)^2) \over D'(D' + 2 \kappa)},}
\eqn\lcII{
R=-\partial_r^2 h = {(-e^{-2 \phi}(2W + W') + 2 D'' (\nabla\phi)^2)
\over (D' + 2\kappa)},}
where $\nabla^2 \phi= \del_r(2  \del_v \phi+ h \del_r \phi)$,
and $ (\nabla \phi)^2 = 2 \del_r \phi \del_v \phi + h (\del_r \phi)^2$.
The remaining local linear combination of the constraint equations is
(conveniently the linear combination needed for matching across the
shock wave),
\eqn\lcIII{\eqalign{
{\it l}^\alpha {\it l}^\beta T_{\alpha \beta} &=
2 D' (\partial_v \phi + \half h \partial_r \phi)^2
+ (\half h \partial_r + \partial_v - \half (\partial_r h) )(\half
\partial_r + \partial_v) D
\cr &+
{\kappa\over 4} ({h\over 2} \partial_r^2 h+
\partial_v \partial_r h -
{1\over 4} (\partial_r h)^2) + t_r(r)}}
where
$t_r$ is determined by the initial state of the quantum vacuum, and for
our collapse, and for static solutions with no net flux at infinity, we
set $t_r$ to  zero in coordinates that are asymptotic to the linear dilaton
vacuum.

For static solutions we put $h = h(r)$, $\phi = \phi(r)$.
With linear dilaton asymptotics at $\phi \to -\infty$, $W \to 4$,
$D \to e^{-2\phi}$, we find to lowest order\foot{Certain choices
of the potentials $D$ and $W$ will give an additional term
in the asymptotics
of $h$  proportional to  $r \exp(-2r)$.},
\eqn\lindilasyms{\eqalign{
h(r) &= 1 - 2 M e^{-2 r}\cr
\phi(r) &= -r~ .}}
One can numerically integrate the static equations using the above as
boundary conditions at $r\to \infty$.
%We carried out the numerical analysis
%for $D = e^{-2\phi} - \gamma^2 e^{2\phi}$ and $W = 4 - \mu^2 e^{4\phi}$.

Now let us consider the behavior of fields near the horizon, $r=r_h$,
of the extremal solution.
We know from the classical analysis in \nsbh, that the classical
extremal solution
has to lowest order $\phi = \phi_0$ and $h = h_0 x^2$ near the horizon,
where  $x = r - r_h$. Let
us look at the quantum corrections to these formulae,
\eqn\phico{\eqalign{
\phi &= \phi_0 + \beta x^{1+\delta}\cr
h &= \alpha_1 x^2 + \alpha_2 x^{3 + \eta}}.}
Plugging this into the static equations we find (evaluating
all functions at $\phi = \phi_0$),
\eqn\pzandhz{\eqalign{
D' W &= \kappa ({W'\over 2} - W)\cr
\alpha_1 &= {e^{-2\phi_0}(W + \half W') \over
D' + 2\kappa} \cr &=
{e^{-2\phi_0}W \over \kappa}\cr
\eta &= \delta \cr
(\delta + 1)(\delta + 2) &= {D''W + D'W - \kappa (\half W'' - W')\over
D'(W + \half W')}.}}
$\beta$ and $\alpha_2$ are related by
\eqn\betalp{
\beta = {-\kappa \alpha_2 (\delta + 2)\over 2\alpha_1 \delta D'}.}

Notice that the solutions have two obvious extensions through $\phi = \phi_0$
(see \sttriv\ for a similar discussion). To study
perturbations by shock waves of matter  we will
restrict to the odd extension which is the smoother of the two continuations
and is the one that is a deformation of the classical geometry,
\eqn\oddext{\eqalign{
\phi &= \phi_0 + \beta x |x|^\delta\cr
h &= \alpha_1 x^2 + \alpha_2 x |x|^{2 + \delta}}~.}
This also means that immediately above the shock wave we will
find two apparent horizons at $\phi_+ < \phi_0$ and at
$\phi_- > \phi_0$, with a geometry qualitatively the same as the
positive mass classical solutions.

We can evaluate the parameters in the expansion near the horizon of
the extremal solution and observe that the
quantum vacuum near the horizon is indeed
a small $\kappa$ deformation of the classical vacuum.
Let us fix
\eqn\pots{
D = e^{-2\phi} - \gamma^2 e^{2\phi}, \qquad {\rm and} \qquad
 W = 4 - \mu^2 e^{4\phi}.
}
For small
$\kappa$ and large $\mu$, we find,
\eqn\params{\eqalign{
e^{-2\phi_0} &= {\mu\over 2} + {\kappa\over 2}\cr
\alpha_1 = 4,\quad \delta & = {2\kappa\over \mu},\quad
{\beta\over\alpha_2} = {1\over 8}}~.}
For definiteness we put $\beta = -1/ \mu$ in the following.

To check that the quantum behavior near $\phi = \phi_0$ that we have
displayed is consistent with the linear dilaton at infinity we
numerically integrated out
from $\phi = \phi_0$ to $\phi = \pm \infty$, and indeed have observed
the linear dilaton vacuum for $\phi \to -\infty$ and the large $\phi$
classical behavior in the other asymptotic regime.

\newsec{\bf The Adiabatic Approximation}

In order to study the semi--classical stability of the
static extremal solutions of the DW models,
consider sending in a
matter shock wave $l^{\mu} l^{\nu} T_{\mu \nu}^f =
2 M \delta(v) $. We define
\eqn\sigdef{
\Sigma = 2 \partial_v \phi + h \partial_r
\phi~,
}
so that the future apparent horizon is the locus of $\Sigma = 0$.
The discontinuity in $\Sigma$ across the shock is
\eqn\dsig{
\delta \Sigma = { {4M } \over { \sqrt{D^{\prime} (D^{\prime}+2 \kappa)}
}}
{}~.}

To obtain explicit expressions we use the potentials
of equation \pots\ and make a large $\mu$, small $\kappa$,
and small $M$ expansion.
Specifically,
we have $ M/ \mu \ll \kappa / \mu \ll 1$.
We may calculate the positions of the
apparent horizons which turn out to be
\eqn\rpos{
r_{\pm}=r_h \pm \sqrt{\mu} ( {M \over { \mu}})^{1 \over {2+\delta}}
}
which is obtained by setting $\Sigma+ \delta \Sigma =0$.
This implies that
\eqn\drshor{
\partial_r \Sigma(r_{\pm} ) = \mp {{ 8} \over {\sqrt{\mu}}}
( {M \over { \mu}} )^{ {1+\delta} \over {2+\delta}}
{}~.}

The discontinuity in $h$ across the shock is given by
\eqn\dhshock{
\delta h = {{8 M} \over {\kappa}} \int dr \bigl(-1 - {{D^{\prime}} \over
{\sqrt{
D^{\prime} (D^{\prime}+2 \kappa) } }} \bigr)~.
}
By inserting this expression into the constraint equation
we find $\partial_v
\Sigma$ at the apparent horizons,
\eqn\dvsig{
\partial_v \Sigma = - {{8M \kappa}\over {\mu^2}}
{}~.}

One may then try to make an adiabatic approximation to compute the
relative positions of the horizons as a function of $v$, by assuming
the equations \drshor\ and \dvsig\
continue to hold for all $v$ if things are changing
slowly enough:
\eqn\dvrpm{
\partial_v {\hat r^{\pm}} = -{\partial_v \Sigma \over
\partial_r \Sigma} =
\mp {{\kappa }\over {\mu}} ({\hat r^{\pm}}
-r_h)
}
which gives
\eqn\rhv{
{\hat r^{\pm}}(v) = r_h + r_0^{\pm} e^{- {{\kappa}\over {\mu}} (v-v_0) }
}
or expressing things in terms of an effective mass, which
measures the difference between the actual mass and the extremal
mass
\eqn\mhv{
M(v) = M_0 e^{-{{2 \kappa}\over {\mu}} (v-v_0) }~.
}
The adiabatic approximation will break down when the energy flux
due to Hawking radiation is comparable to $M(v)$, so
will be valid as long as $\kappa / \mu \ll 1$.

These expressions are almost identical to ones obtained in
\refs{ \sttriv} in the  case of the spherically
symmetric reduction of the Reissner--Nordstrom black hole,
apart  from slight changes in numerical coefficients.
The RN case is obtained by picking different potentials $D$
and $W$. In this case, the adiabatic approximation is valid as
long as $\kappa / Q^2 \ll 1$, where $Q$ is the charge of the
extremal black hole.

\newsec{\bf The Numerical Collapse}

When an extremal solution is perturbed by an incoming flux
of matter the picture one expects is the following:
above the shock wave two apparent
horizons form on either side of the  line $\phi_0$, and as the
black hole evaporates these horizons approach each other,
with the solution settling back down to the extremal solution
at $\scri^+_R$.
In this section we describe numerical solutions for a shock
wave of in--falling matter impinging upon the extremal solution
of the DW model described in section 3. We also consider
the analogous calculation for the RN type model. Our aim
is to test whether the extremal black hole is stable,
and in what manner the evaporation proceeds.

Another approach would be to do a linear stability analysis of the
problem, where one might hope to be able to understand things
analytically. In fact, it seems this is not so for the
back--reaction corrected equations. A numerical analysis
of this problem is tractable, but is of comparable
difficulty to numerically solving the full nonlinear equations,
so we choose to do the latter.

In the context of the CGHS model, numerical results
have previously been obtained in \refs{\hastew, \dlowe, \pist} .
Here we are solving equations of precisely the same form,
but with somewhat different potentials. See \refs{\dlowe}
for a description of the numerical algorithm used in this paper,
and also appendices A and B, where the gauge and coordinate
choices, and equations of motion
are stated. For numerical purposes it is convenient to
work on a grid of null lines, so conformal gauge is appropriate.
This should be borne in mind when making quantitative comparisons
with the results of the previous section, where it was necessary
to use light--cone gauge to get analytical results.
\ifig\ftwo{Plot of motion of apparent horizons for DW model,
in the strong back--reaction regime.
The lower line is the outer apparent horizon, the middle line
is $\phi=\phi_0$, and the upper line is the inner apparent
horizon.}
{\epsfysize=3in\epsfbox[150 50 612 600]{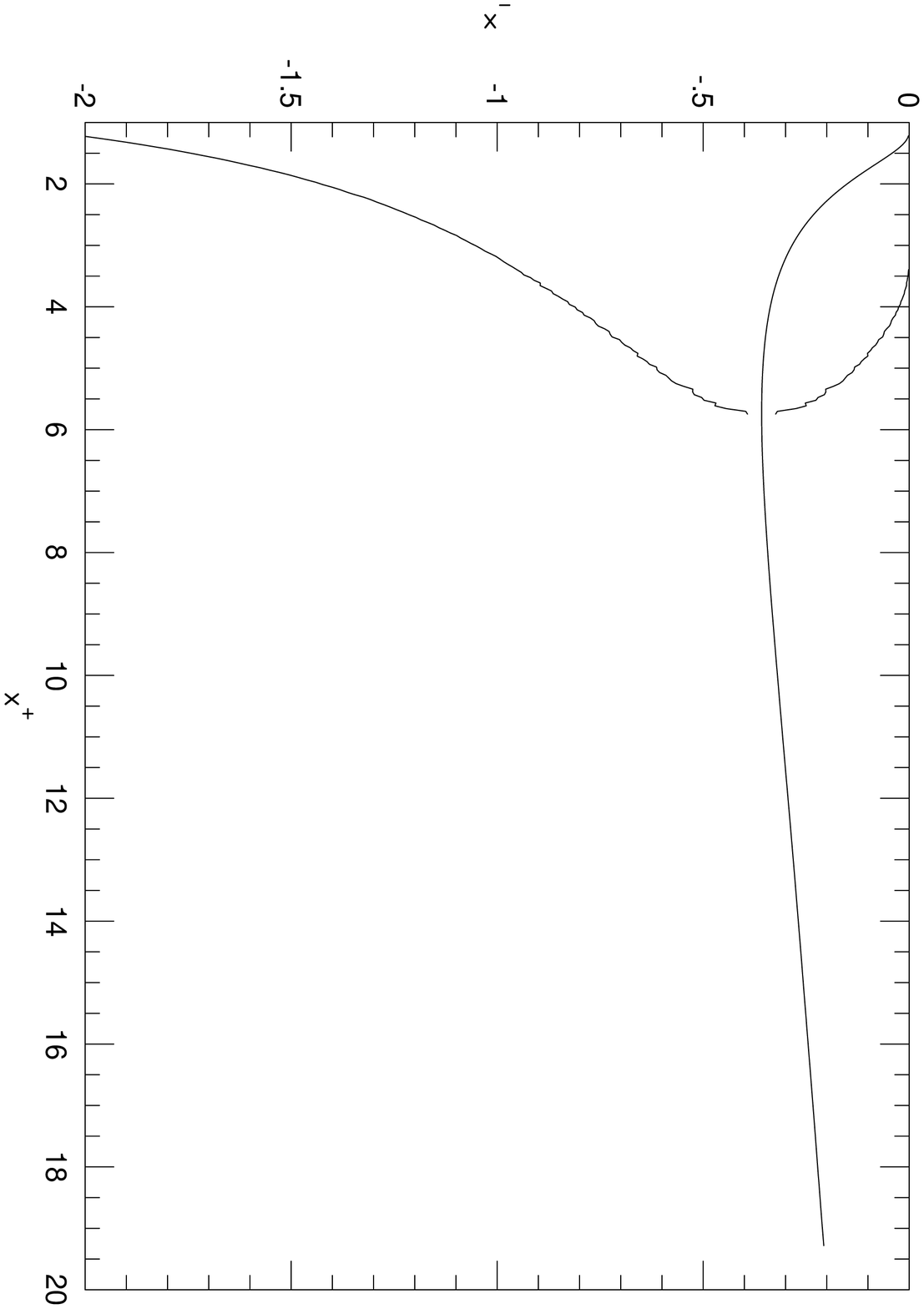}}

For the DW model of equation \pots\ , with $\kappa = 10$,
$\mu=15$, $\gamma =8$,
and shock mass $M_s = 1.5$, the results are plotted in
\ftwo\ . Note that here $\kappa / \mu >1$ so one is in the
strong back--reaction regime and
the adiabatic approximation of the previous
section is not expected to hold.
The integration is stopped near $x^-=0$ where the line $\phi=\phi_0$
starts out. This means these calculations will hold for
either of the extensions described in the previous section.
The horizons meet at finite
proper time, and the line $\phi_0$ goes from being spacelike
to timelike at this point.
Note the equations of motion imply that $\del_+ \del_- \phi =0$
when an apparent horizon ( where $\del_+ \phi=0$ )
intersects $\phi=\phi_0$.
The position of the
apparent horizons $\hat x^-(x^+)$ may be found by solving
\eqn\hormo{
{{\del {\hat x}^- }\over {\del x^+}} = - {{\del_+^2 \phi}\over {\del_-\del_+
\phi}}
}
which indicates that ${\del {\hat x}^- }/ {\del x^+} $ blows up
as $\del_+ \del_- \phi \to 0$,
as long as $\del_+^2 \phi$ remains finite.
This is precisely what is happening
at the meeting point of the apparent horizons, consistent with
the numerical results. Also note that the wiggles in the path of
the outer horizon are due to $\del_- \del_+ \phi$ becoming very small
in this region. This causes the error in the path of the line
$\del_+ \phi=0$ to be much larger than, say, the error
in the path of a line of constant $\phi$. The conclusive
indication that the horizons meet comes from the observation
that the line $\phi=\phi_0$ becomes timelike. Numerical convergence
has been checked by varying the stepsize, and varying the position
of the initial surface.

\ifig\fthree{Plot of the curvature along a line of constant
$\phi=-1.2$, for the DW model, in the strong
back--reaction regime.
Here the range of $x^+$ extends much further out.
The curvature approaches a constant value which is the same as
the initial curvature of the extremal solution.}
{\epsfysize=3in\epsfbox[150 50 612 600]{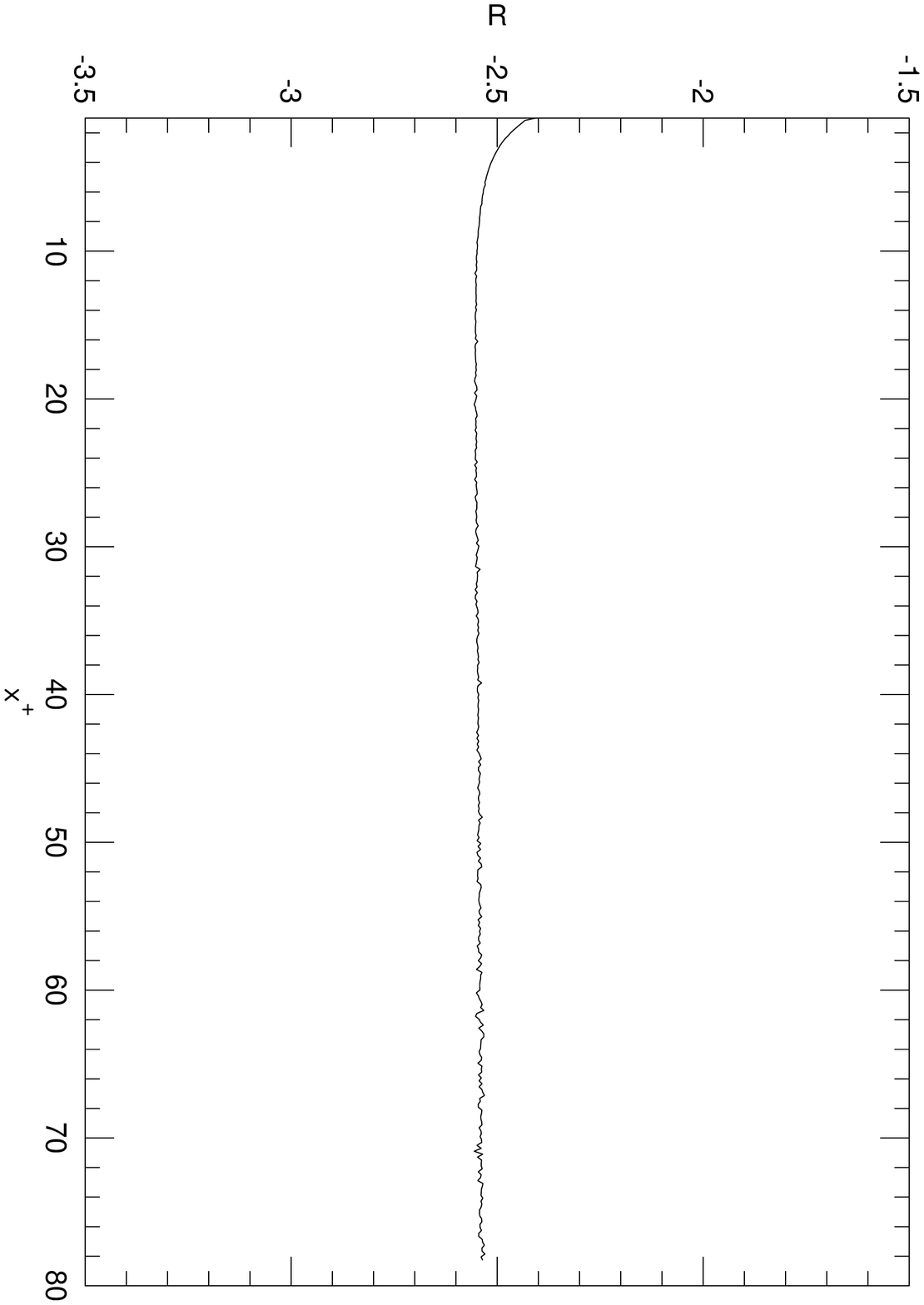}}

Following a line of constant $\phi$ as $x^+$ becomes large,
one finds the curvature approaches a constant which equals
the initial curvature of the extremal solution, as shown
in  \fthree\ . This indicates that despite the
fact that the apparent horizons have met, the solution
still settles back down to the zero temperature extremal state.

\ifig\ffour{Plot of the path of a line of constant $\phi=-1.2$,
for the DW model, in the strong
back--reaction regime.
As $x^+$ becomes large the line appears to asymptotically
approach a null line.}
{\epsfysize=3in\epsfbox[150 50 612 600]{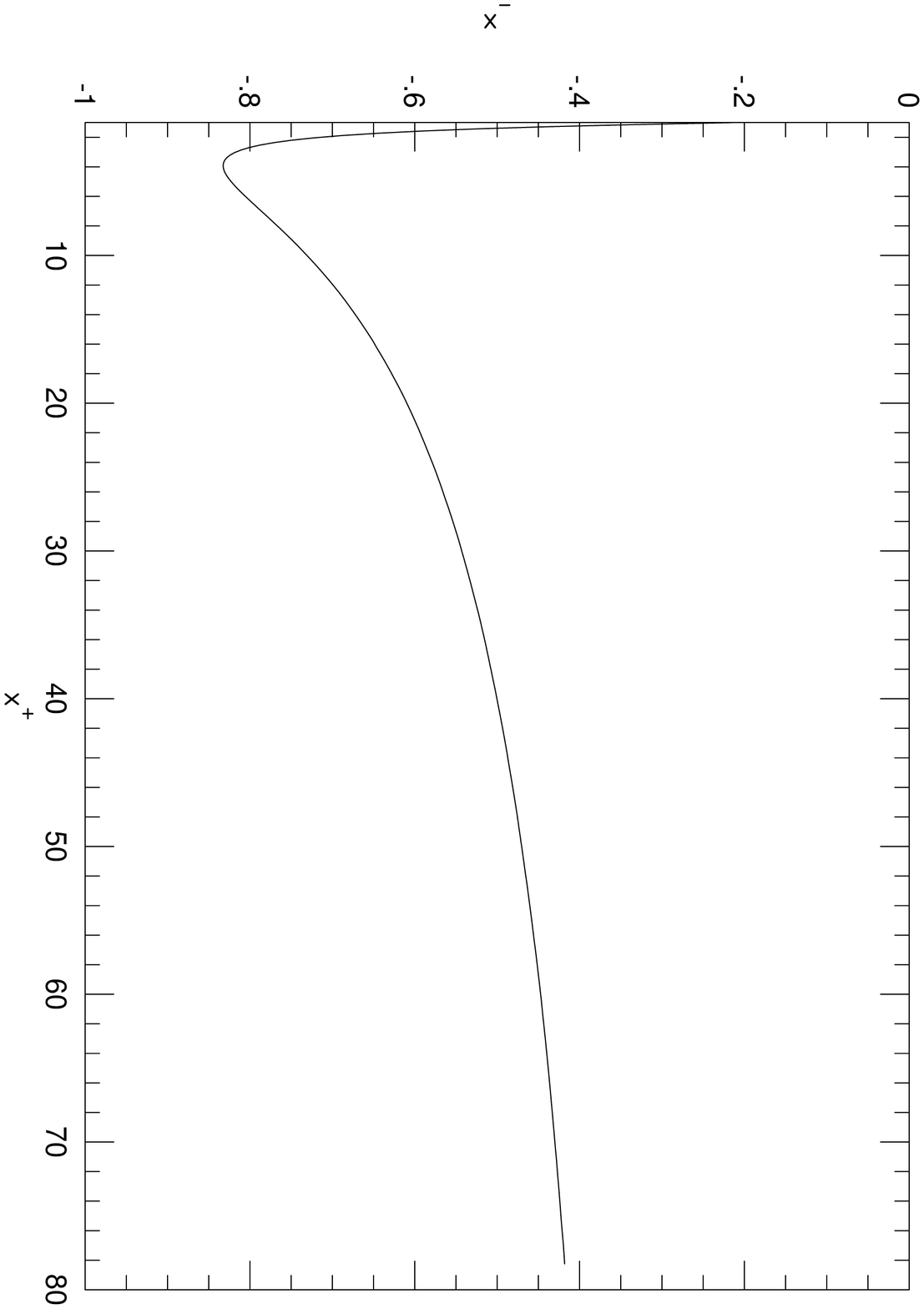}}
Lines of constant $\phi$ appear to approach a null line $x^-=x^-_0$,
which becomes a global horizon,
as shown in  \ffour\ . It is difficult to tell from the numerics
whether $x^-_0=0$, or whether it is shifted out to more negative
$x^-$ by the in--falling matter. More sophisticated numerical
calculations are needed to answer this question definitively.

Simulations were also performed for the weak back--reaction regime
where
$\kappa / \mu \ll 1$, when the adiabatic approximation
is expected to hold. Here potential numerical errors were
somewhat larger, but the results were found to be consistent
with the adiabatic approximation. The apparent horizons
persisted for very large $x^+$, approaching the critical line
$\phi= \phi_0$.

\ifig\ffive{Plot of motion of apparent horizons for RN model,
in the strong back--reaction regime. The
lower line is the outer apparent horizon, the middle line
is $\phi=\phi_0$, and the upper line is the inner
apparent horizon.}
{\epsfysize=3in\epsfbox[150 50 612 600]{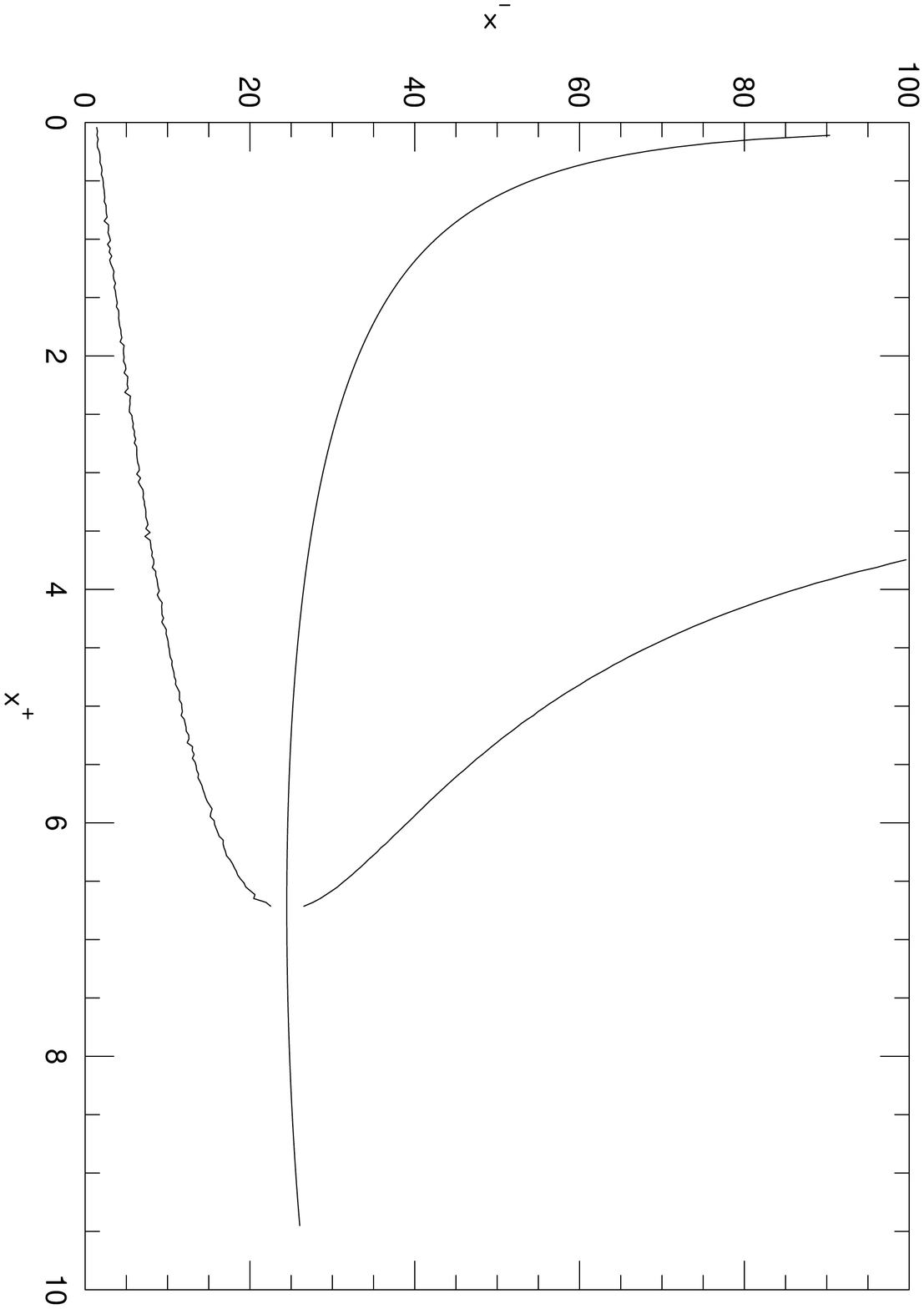}}

Qualitatively similar results are obtained for the
case of the RN model. In the strong back--reaction regime
where $\kappa / Q^2 > 1$, with
$\kappa = 200$, $Q^2 = 60$, and shock
mass $M_s= .08$, results are shown in  \ffive\ . One difference
here is that coordinates are chosen so the line $\phi=\phi_0$
starts out at $x^- = +\infty$. Again the solution appears
to settle back down to the extremal solution as $x^+$ becomes large,
after the apparent horizons have met.
In the weak back--reaction regime ($\kappa / Q^2 \ll 1$),
results consistent
with the adiabatic approximation were found, with the apparent
horizons approaching the the critical line out to large
values of $x^+$.

\newsec{\bf Discussion and Conclusions}

\ifig\fsix{Penrose diagrams showing the end--points of the black
hole evaporation. The case of  strong back--reaction is
shown on the left, the case of weak back--reaction is shown on the
right.}
{\epsfysize=3in\epsfbox{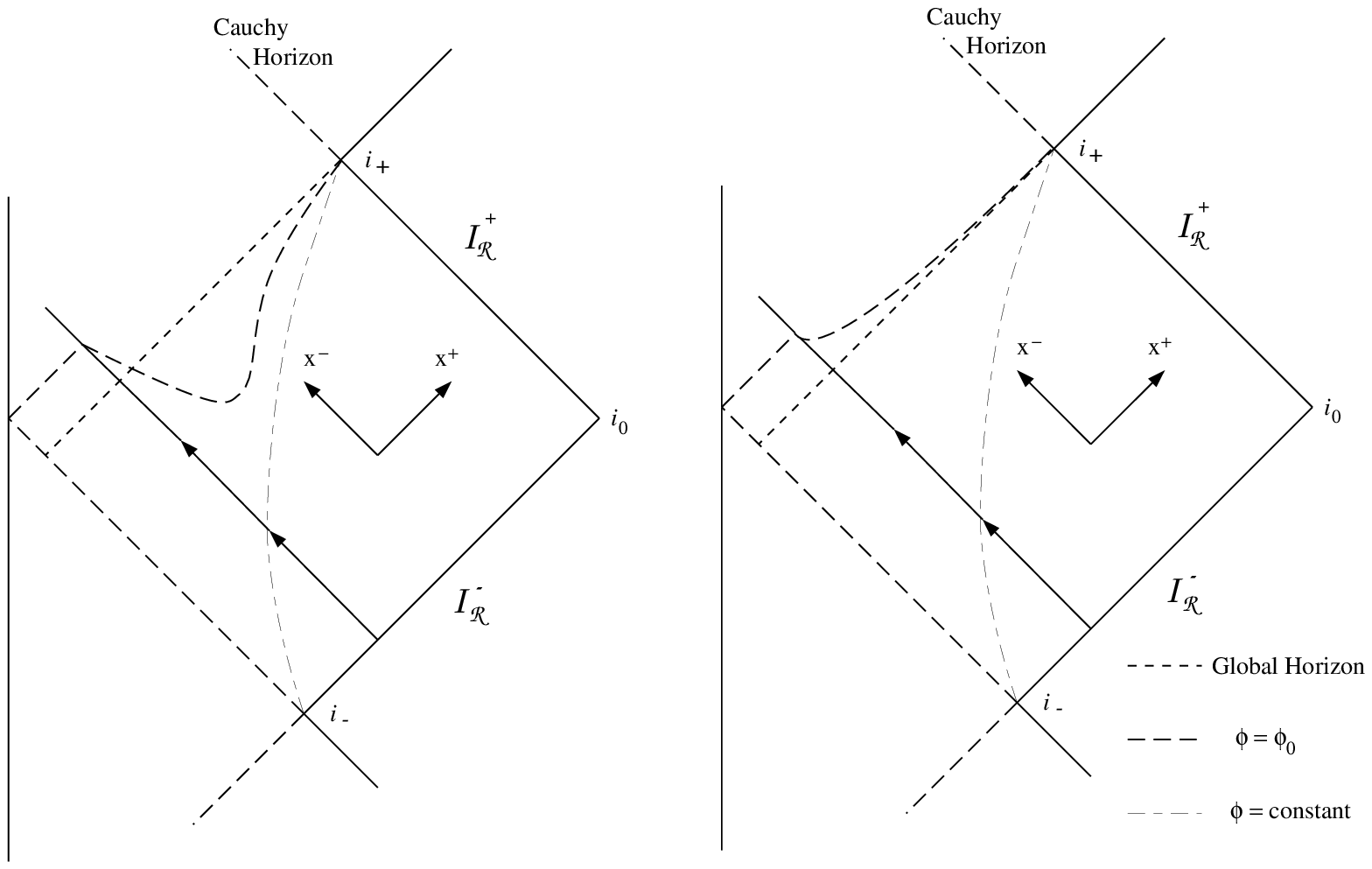}}

The picture of the end--point of the evaporation of these
two--dimensional extremal black holes which is consistent with
the numerical calculations
is shown in
\fsix\ . In the strong back--reaction regime, lines of constant
$\phi$ approach a global horizon,
and the solution settles back down to the extremal one.
After the apparent horizons meet the line $\phi = \phi_0$ turns timelike
and becomes asymptotically null as it approaches the global horizon.
In the weak back--reaction regime the apparent horizons
meet at $\scri^+_R$, and the critical line $\phi= \phi_0$
remains spacelike.

Our results indicate the presence
of two qualitatively different regimes separated by
some critical value of $\kappa/ \mu$ in the DW case, and
$\kappa / Q^2$ in the RN case. This should not be
too surprising, since a similar phenomena occurs in
the case of the damped harmonic oscillator, where there is a
critical value of the damping separating two
regimes of behavior. Although we have studied two
particular models, we believe that this behavior will be
generic to the class of models in which $W(\phi)$ has a simple
zero, which includes the nonsingular models discussed in
section 2.

Because for large $\kappa/ \mu$ the apparent horizons collide,
locally, the end--point of the evaporation of these extremal
solutions looks like an $M^2 < Q^2$ static solution in the
case of the RN model. One may wonder if some two--dimensional
quantum positive
mass theorem prevents such an unexpected occurrence. However,
positive mass theorems only give us information about the asymptotic
structure of the space--time and certainly do not preclude the type of
relaxation that we observe here. Classically the DW and RN models obey
positive mass theorems \parkst\ with the mass defined on space--like surfaces
that become asymptotically null along ${\cal I}_R^+$ and have their left
boundary at the meeting of the two global horizons of the extremal
solution below the shock wave. Unfortunately the quantum positive mass
theorems \bilal\ do not appear to give any useful information in their present
formulation due to the quantum corrections to the vacuum. Further,
quantum positive mass theorems  require a nonsingular spacelike
surface that is asymptotically linear dilaton vacuum at both ends, see
\refs{\bilal,\bilkog,\dealwis}. In our situation this of course can
never be the
case, since the left boundary of our spacelike hypersurfaces meets the
horizon of the
extremal solution which looks distinctly not like the LDV.

These results appear to confirm the conjecture that zero temperature
extremal (in the sense of RN) two--dimensional
black holes are semi--classically stable. In the case of the DW models
the evaporation of nonextremal black holes proceeds in a completely
nonsingular way,
realizing the original objectives of CGHS.
Information loss is inevitable in these models at the
semi--classical level\foot{See
\refs{\nsbh,\sttriv,\info}\ for related discussions.},
since an in--falling flux
of matter will always produce correlations with the region behind the
global horizon, and hence inaccessible to an observer who finds
themselves
at $i_+$ after an infinite proper time. The information loss is of a
somewhat benign type though. For observers outside the global horizon
the quantum mechanics that they participate in is unitary, indeed the
spacetime before the Cauchy horizon can (by definition) be foliated by
space--like Cauchy hypersurfaces. The information that is no longer
accessible to them is stored in
a stable remnant, one of an infinitely degenerate set of possible
final states. This infinity of states corresponds to all the possible
field configurations on a space--like slice through region III (\fone ).
These remnants avoid the problems of overproduction in
external fields and divergences in virtual loops by nature of their
large internal geometry \info\ .

Let us note that the above discussion is based on
semi--classical reasoning.
It is still possible that when
space--time is properly second quantized the information
loss problem will be cured. Highly nonlocal quantum gravity
effects may wind up giving the remnants a very long, but finite lifetime.
If this lifetime is of order the age of the universe,
experimenters making measurements over shorter times
will still see an effective loss of information, which
they would attribute to the existence of ``stable'' remnants.

The conjectured extremal state possesses a Cauchy
horizon as shown in \fone .
We would like to comment on the sense in which this horizon is
traversable\foot{See \chhart\  and references
therein for earlier discussions.} and the possibility of information
loss for observers who traverse it.  We should first
note that the Cauchy horizon in our model appears to be
a double horizon, and as
discussed in \triv\ has a softened divergence of the stress tensor as
compared to the inner horizon of a nonextremal space--time. It is
also possible that the disturbance produced by the shock wave will
separate the local horizon and the global Cauchy horizon in a manner
similar to the mass inflation models of Poisson and Israel \PI . In mass
inflation it was shown that an infinite tidal force appears along the
Cauchy horizon, but the singularity is weak
\ori\ in the sense that an observer can traverse it without
getting stretched infinitely.

Let us assume then that a traversable Cauchy horizon is present. An
observer who passes into the
region above this Cauchy horizon faces  a
potential loss of unitarity associated with the lack of a global
foliation of the space--time by space--like Cauchy hypersurfaces.
%One (albeit speculative) way out of information loss for such
%an observer would be a modified version of Strominger's conveyer belt
%mechanism \ref\stconv{A. Strominger, unpublished.}. In our case the
%conveyer belt would operate
%along ${\cal I}_R^+$, carrying information to $i_+$ for evolution past
%the Cauchy horizon. This would not involve any acausality.
The existence of a Cauchy horizon, it's nature if it exists
and the necessity of a unitarity
restoring mechanism for those who cross it, are the subject of our
ongoing investigations. The answers to these questions should shed light
on an S--matrix description of quantum gravity in the presence of
remnant--like objects.

\bigskip
\centerline{{\bf Acknowledgments}}

The research of D.L. was supported in part by
DOE grant DE-AC02-76WRO3072, NSF grant PHY-9157482, and James
S. McDonnell Foundation grant No. 91-48.
D.L. would like to thank C. Callan and A. Bilal for useful
discussions.
The research of M.O'L. was
supported in part by the Department of Energy under grant
No.DE-FG05-90ER40559 and the National Science Foundation under Grant No.
PHY89-04035. M.O'L. would like to thank T. Banks for innumerable
enlightening conversations on the subject, and M. Douglas, S. Shenker
and A. Strominger for conversations. He would also like to thank the ITP
at Santa Barbara for hospitality during part of this work.
We thank A. Strominger and S. Trivedi for comments on a previous
draft of this manuscript.

\appendix{A}{Numerical method for DW model}

The numerics are performed in conformal gauge
\eqn\conf{
g_{++}=g_{--}=0, \qquad g_{+-} = g_{-+} = -\half e^{2 \rho}~.
}
The linear dilaton vacuum in the coordinates we choose is
\eqn\lindil{
\phi = \rho = -\half \log( -x^+ x^-)
}
while a solution with mass $M$ looks like
\eqn\masssol{
\phi=\rho = -\half \log(M-x^+ x^-)
}
in the limit $x^+ x^- \to -\infty$. The equations to be
solved are
\eqn\emotion{
\eqalign{
\del_+ \del_- \rho &=
{{ - D^{\prime \prime} \del_+ \phi \del_- \phi - {1\over 4} e^{2 \rho-
2\phi} (W+ {{W^{\prime} }\over 2} ) } \over { D^{\prime} + 2\kappa}} \cr
\del_+ \del_- \phi &=
- {{D^{\prime \prime} \del_+\phi \del_- \phi + {1\over 4} e^{2 \rho-
2\phi} W + \kappa \del_+ \del_- \rho } \over {D^{\prime}} }~. \cr }
}
Here $D= e^{-2 \phi} - \gamma^2 e^{2 \phi} $ and $W= 4 - \mu^2 e^{4
\phi}$.
The boundary conditions are that along $x^+ =1$ the solution
correspond to the extremal solution discussed in section 3.
Above the shock wave along $\scri^-$ the solution should agree
with the classical shock solution. This amounts to setting
\eqn\dpphi{
\del_+ \phi = \del_+ \rho \sim
{ {x^-+M} \over {2(M-x^+ x^- - M x^+ )} }
}
as $x^- \to -\infty$. In practice a large negative initial
value of $x^-$ is chosen.

\appendix{B}{Numerical method for the RN model}

Here the numerics are also performed in conformal gauge
\conf\ . Asymptotically as $x^+-x^- \to \infty$
the solutions approach the vacuum
\eqn\rnvac{
\phi = - \log( \half(x^+ -x^-) ), \qquad \rho=0~.
}
The equations to be solved are
\eqn\rnmotion{
\eqalign{
\del_+ \del_- \rho &= { {\del_+ \phi \del_- \phi
+ {1\over 4} e^{2 \rho} (e^{2\phi} - 2 Q^2 e^{4 \phi} ) }\over
{ 1 - {{\kappa} \over 2} e^{2 \phi} } } \cr
\del_+ \del_- \phi &= \del_+ \del_- \rho + \del_+ \phi \del_- \phi
+ {{Q^2} \over 4} e^{2 \rho+4\phi} ~.\cr }
}
Here the boundary conditions are that along $x^+=0$ the solution
match onto the quantum corrected extremal solution \refs{\triv},
while above the shock wave along $\scri^-$ the solution agree
with a classical shock solution. This means
\eqn\dprnphi{
\del_+ \phi \sim -{ 1\over {x^+ -x^-}} + {{4M} \over {(x^+-x^-)^2} }
\bigl( 1 -\log( \half(x^+-x^-)) \bigr), \qquad \del_+ \rho \sim
{ {2M} \over {(x^+-x^-)^2} }
}
as $x^- \to -\infty$.

\vfill\eject
\listrefs
\end